# Reply to the Comment on 'Unraveling Photoinduced Spin Dynamics in the Topological Insulator Bi$_2$Se$_3$'

M. C. Wang,[1,2] S. Qiao,[3,4] Z. Jiang,[5] S. N. Luo,[1,2] and J. Qi[1,2]

[1]The Peac Institute of Multiscale Sciences, Sichuan 610031, China

[2]Key Laboratory of Advanced Technologies of Materials, Southwest Jiaotong University, Sichuan 610031, China

[3]State Key Laboratory of Functional Materials for Informatics, Shanghai Institute of Microsystem and Information Technology,
Chinese Academy of Sciences, Shanghai 200050, China

[4]School of Physical Science and Technology, ShanghaiTech University, Shanghai 200031, China

[5]School of Physics, Georgia Institute of Technology, Atlanta, Georgia 30332, USA

In the Comment [1], the author argues that in our work [2], both electron optical phonon interaction timescales for $A_{1g}^1$ mode (~300-700 fs) and $E_g^2$ mode (~30 fs) are too short in the photoexcited Bi$_2$Se$_3$ single crystals. Instead, the Comment author contends that the electron optical phonon scattering process has a timescale larger than 1 ps.

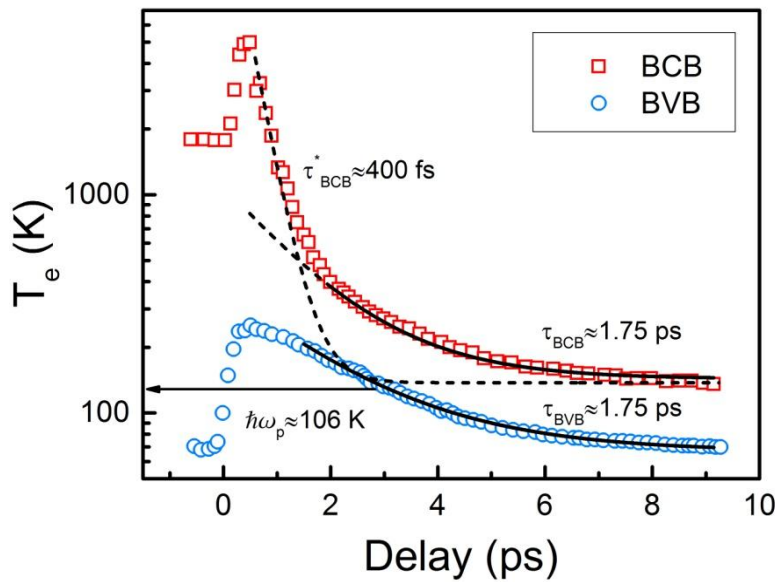

Figure C1  The transient electronic temperatures in Bi$_2$Se$_3$ measured at 70 K (adapted from Ref. [4]). The black arrow indicates the phonon temperature $k_B T_p$ ~106 K ($1.5 k_B T_e$ ~71 K) for the $A_{1g}^1$ phonon ($\hbar\omega_p$ ~2.2 THz ≈ 9.1 meV).

First, the electron phonon scattering times, associated with the dominant $A_{1g}^1$ optical phonons and low energy acoustic phonons, can be unambiguously revealed by comparing the bulk dominated ΔR/R with the tr-ARPES experiments from two different groups (Gedik [3] and Shen [4]). Fig. C1 shows the corresponding data, where the decay of the transient electronic temperature ($T_e^{BCB}$) of the bulk conduction band (BCB) in $Bi_2Se_3$ is well consistent with the experiment in Ref. [4]. $T_e^{BCB}$ can be fit well using two exponential decays with distinct relaxation time (dashed lines in Fig. C1). Specifically, $\tau^*_{BCB} \approx 400$ fs and $\tau_{BCB} \approx 1.75$ ps are obtained, suggesting that the hot carriers in BCB undergo two successive energy relaxations. These two timescales are also clearly observed in our ΔR/R measurements [2]. Evidently, cooling of bulk valence band (BVB) and BCB characterized by $\tau_{BVB}$ and $\tau_{BCB}$ is dominated by the acoustic phonon-mediated cooling, because (1) fitting the data for $T_e^{BVB}$ <71 K ($T_p(A_{1g}^1)$~106 K) or $T_e^{BVB}$ <250 K nearly arrives the same $\tau_{BVB}$ (~1.75 ps), (2) no emission of the dominant $A_{1g}^1$ optical phonons for $T_e^{BVB}$<71 K ($T_p(A_{1g}^1)$~9.1 meV~106 K), and (3) $\tau_{BCB} \approx \tau_{BVB}$. Therefore, the initial fast electron cooling in BCB characterized by $\tau^*_{BCB} \approx 400$ fs can only be ascribed to the electron hot optical phonon scattering processes, where the $A_{1g}^1$ phonon is dominant. Thus, the electron $A_{1g}^1$ optical phonon interaction time is <1 ps, and the successive picosecond relaxation process is associated with the acoustic phonons. The authors in Ref. [3] also arrive the similar conclusions.

Second, the rarely explored ultrafast electron phonon interaction time for the weak $E_g^2$ mode has already been observed and discussed recently in several works [5,6]. In fact, the coherent phonons at THz frequencies have been extensively studied for almost 30 years [5-20]. A unifying mechanism for phonon generation in both transparent and opaque materials was first presented by Merlin et al. [5,15-16]. Phenomenologically, the lattice motion is described by $d^2Q/dt^2 + \Omega^2 Q = \vec{F}(t)$, where the lifetime of driving force $\vec{F}(t)$ in opaque materials reflects the electron phonon interaction timescale. In Bi and Sb, with structures having the same point group $3\bar{m}$ as that of $Bi_2Se_3$, the electron phonon interaction time for $E_g^2$ mode is experimentally found to be ~2-13 fs [5], while the long lifetime of driving force for $A_{1g}^1$ mode is characterized by the decay of photoexcited carriers, i.e. the decay of electron temperature discussed above. Such short timescale associated with the $E_g^2$ mode, breaking the three-fold rotational symmetry, is further confirmed by first-principles DFT calculations [6], where the ultrafast scattering is explained by rapid decay of the low-symmetry component of the photoexcited carrier distribution and of the corresponding low symmetry atomic forces [5,6]. Therefore, the electron $E_g^2$ optical phonon interaction timescale of ~30 fs found in our experiment is fully consistent with previous experiments and theory.

Finally, the Comment author states that the transient reflectivity signal can be associated with the photo-Dember effect or the optical carrier "shock wave" injection process. Since they are used to phenomenologically explain $F(t)$ [9,12,14], and have already be fully accounted by the above model and calculations [5,6,15-16], they do not conflict with our findings that the electron optical phonon interaction has timescales of ~30 fs and 300-700 fs for $E_g^2$ and $A_{1g}^1$ modes.